\documentclass[a4paper, twoside, reqno, 12pt, dvips]{amsart}
\RequirePackage{fix-cm} 

\usepackage{fixltx2e}     

\usepackage[english]{babel}
\usepackage[latin1]{inputenc}

\usepackage{eucal}
\usepackage{esint}
\usepackage{amsgen}
\usepackage{amsthm}
\usepackage{xspace}
\usepackage{amssymb}
\usepackage{amsmath}
\usepackage{upgreek}
\usepackage{MnSymbol}
\usepackage{amsfonts}
\usepackage{verbatim}
\usepackage{mathrsfs, dsfont}

\usepackage[nice]{nicefrac}

\usepackage{a4wide}

\usepackage{microtype}
\usepackage[bf, up, hang]{caption}

\usepackage{indentfirst}
\usepackage{graphicx}
\usepackage{subfigure}
\usepackage[section]{placeins}
\usepackage{psfrag}

\usepackage[usenames, dvipsnames, pdf]{pstricks}
\usepackage{epsfig}
\usepackage{pst-grad} 
\usepackage{pst-plot} 

\usepackage{color}
\definecolor{oneblue}{rgb}{0.0, 0.0, 0.85}
\definecolor{darkgrey}{rgb}{0.273, 0.281, 0.30}

\usepackage[colorlinks,
            urlcolor=oneblue,
            linkcolor=oneblue,
            citecolor=oneblue,
            bookmarksopen=false,
            pagebackref]{hyperref}

\usepackage[compact]{titlesec}

\titleformat{\section}{\normalfont\Large\bfseries\sffamily\center\color{darkgrey}}{\thesection.}{0.5em}{}{}
\titleformat{\subsection}{\normalfont\large\bfseries\sffamily\color{darkgrey}}{\thesubsection.}{0.4em}{}{}
\titleformat{\subsubsection}{\normalfont\normalsize\bfseries\sffamily\color{darkgrey}}{\thesubsubsection.}{0.3em}{}{}

\titlespacing*{\section}{1.0em}{1.0em}{0.8em}[0em]
\titlespacing*{\subsection}{1.0em}{1.0em}{0.8em}[0em]
\titlespacing*{\subsubsection}{1.0em}{0.7em}{0.6em}[0em]

\usepackage{titletoc}

\contentsmargin{0.0em}
\dottedcontents{section}[2.5em]{\addvspace{0.45em}\bfseries}{1.0em}{0pc}
\dottedcontents{subsection}[4.5em]{}{2.6em}{0pc}
\dottedcontents{subsubsection}[5.5em]{}{3.0em}{0pc}

\usepackage{fancyhdr}
\usepackage{lastpage}

\newcommand*\Title{Extreme wave run-up on a vertical cliff}
\newcommand*\Authors{F.~Carbone, D.~Dutykh et al.}

\usepackage{acronym}
\acrodef{SGN}[SGN]{Serre--Green--Naghdi}

\numberwithin{equation}{section}

\renewcommand{\a}{\upalpha}
\newcommand{\R}{\mathcal{R}}
\newcommand{\m}{$\mathsf{m}$}
\newcommand{\s}{$\mathsf{s}$}
\renewcommand{\H}{\mathcal{H}}
\renewcommand{\O}{\mathcal{O}}

\newcommand{\half}{{\textstyle{1\over2}}}
\newcommand{\third}{{\textstyle{1\over3}}}

\begin{document}

\title[\Title]{Extreme wave run-up on a vertical cliff}

\author[F.~Carbone]{Francesco Carbone}
\address{University College Dublin, School of Mathematical Sciences, Belfield, Dublin 4, Ireland}
\email{Francesco.Carbone@ucd.ie}

\author[D.~Dutykh]{Denys Dutykh$^*$}
\address{University College Dublin, School of Mathematical Sciences, Belfield, Dublin 4, Ireland \and LAMA, UMR 5127 CNRS, Universit\'e de Savoie, Campus Scientifique, 73376 Le Bourget-du-Lac Cedex, France}
\email{Denys.Dutykh@ucd.ie}
\urladdr{http://www.denys-dutykh.com/}
\thanks{$^*$ Corresponding author}

\author[J.~Dudley]{John M. Dudley}
\address{D\'epartement d'Optique P.M.~Duffieux, Universit\'e de Franche-Comt\'e, Institut FEMTO-ST CNRS UMS 6174, Besan\c{c}on, France}
\email{John.Dudley@univ-fcomte.fr}

\author[F.~Dias]{Fr\'ed\'eric Dias}
\address{CMLA, ENS Cachan, CNRS, 61 Avenue du Pr\'esident Wilson, F-94230 Cachan, France \and University College Dublin, School of Mathematical Sciences, Belfield, Dublin 4, Ireland}
\email{Frederic.Dias@ucd.ie}

\begin{abstract}
Wave impact and run-up onto vertical obstacles are among the most important phenomena which must be taken into account in the design of coastal structures. From linear wave theory, we know that the wave amplitude on a vertical wall is twice the incident wave amplitude with weakly nonlinear theories bringing small corrections to this result. In this present study, however, we show that certain simple wave groups may produce much higher run-ups than previously predicted, with particular incident wave frequencies resulting in run up heights exceeding the initial wave amplitude by a factor of 5, suggesting that the notion of the design wave used in coastal structure design may need to be revisited.  The results presented in this study can be considered as a note of caution for practitioners, on one side, and as a challenging novel material for theoreticians who work in the field of extreme wave - coastal structure interaction.

\bigskip
\noindent \textbf{\keywordsname:} run-up; wave/wall interaction; Serre equations; coastal structures; wave groups

\end{abstract}

\maketitle
\tableofcontents
\thispagestyle{empty}

\section{Introduction}

The robust design of various coastal structures (such as sea-walls, breakwaters, etc.) relies on the accurate estimation of the wave loading forces. To this end, engineers have introduced the notion of the so-called \emph{design wave}. Once the particular characteristics of this design wave are determined, the pressure field inside the bulk of fluid is usually reconstructed (in the engineering practice) using the \cite{Sainflou1928} or \cite{Goda2010} semi-empirical formulas. However, there is a difficulty in determining the wave height to be used in design works. Sometimes, it is taken as the significant wave height $H_{\nicefrac{1}{3}}$, but in other cases it is  $H_{\nicefrac{1}{10}}$ (the average of 10\% highest waves) that are substituted into the wave pressure formulas.  If we take, for example, an idealized sea state which consists only of a single monochromatic wave component with amplitude $a_0$, its wave height $H_0$ can be trivially computed
\begin{equation}
  H_0 \equiv H_{\nicefrac{1}{3}} \equiv H_{\nicefrac{1}{10}} = 2a_0.
\end{equation}
Consequently, the design wave will have also the height equal to $2a_0$.

In the present study, we show that even such simple monochromatic sea states, subject to the nonlinear dynamics over a constant bottom, can produce much higher amplitudes on a vertical wall. Namely, we show below that some wave frequencies can lead to an extreme run-up of the order of $\approx 5.5a_0$ on the cliff. The results presented in this study suggest that the notion of the design wave has to be revisited. Moreover, the mechanism elucidated in this work can shed some light onto the freak wave phenomenon in the shallow water regime, where we recall in this context that over 80\% of reported past freak wave events have been in shallow waters or coastal areas \cite{Nikolkina2011, O'Brien2013}.

It is well known that wave propagation on the free surface of an incompressible
homogeneous inviscid fluid is described by the Euler equations combined with nonlinear boundary conditions on the free surface \cite{Stoker1957}. However, this problem is difficult to solve over large domains and, consequently, simplified models are often used. In particular, in this study we focus our attention on long wave propagation. A complete description of wave processes, including collisions and reflections, is achieved by employing two-way propagation models of Boussinesq type \cite{BC}. Taking into account the fact that we are interested here in modeling (potentially) high amplitude waves, we adopt the fully nonlinear \acf{SGN} equations \cite{Serre1953, Green1974, Green1976, Zheleznyak1985}, which make no restriction on the wave amplitude. Only the weak dispersion assumption is adopted in the mathematical derivations of this model \cite{Wei1995, Lannes2009, Dias2010}.

\begin{figure}
\centering
\subfigure[]{\includegraphics[width=0.69\textwidth]{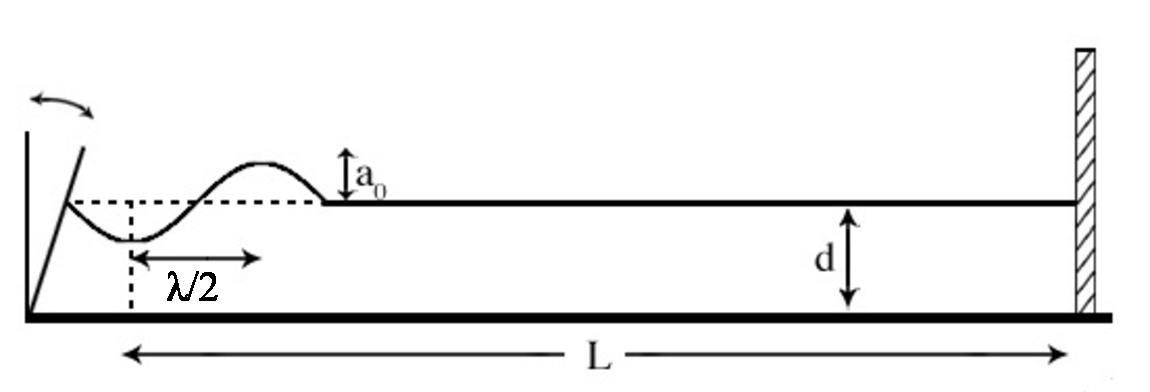}}
\subfigure[]{\includegraphics[width=0.69\textwidth]{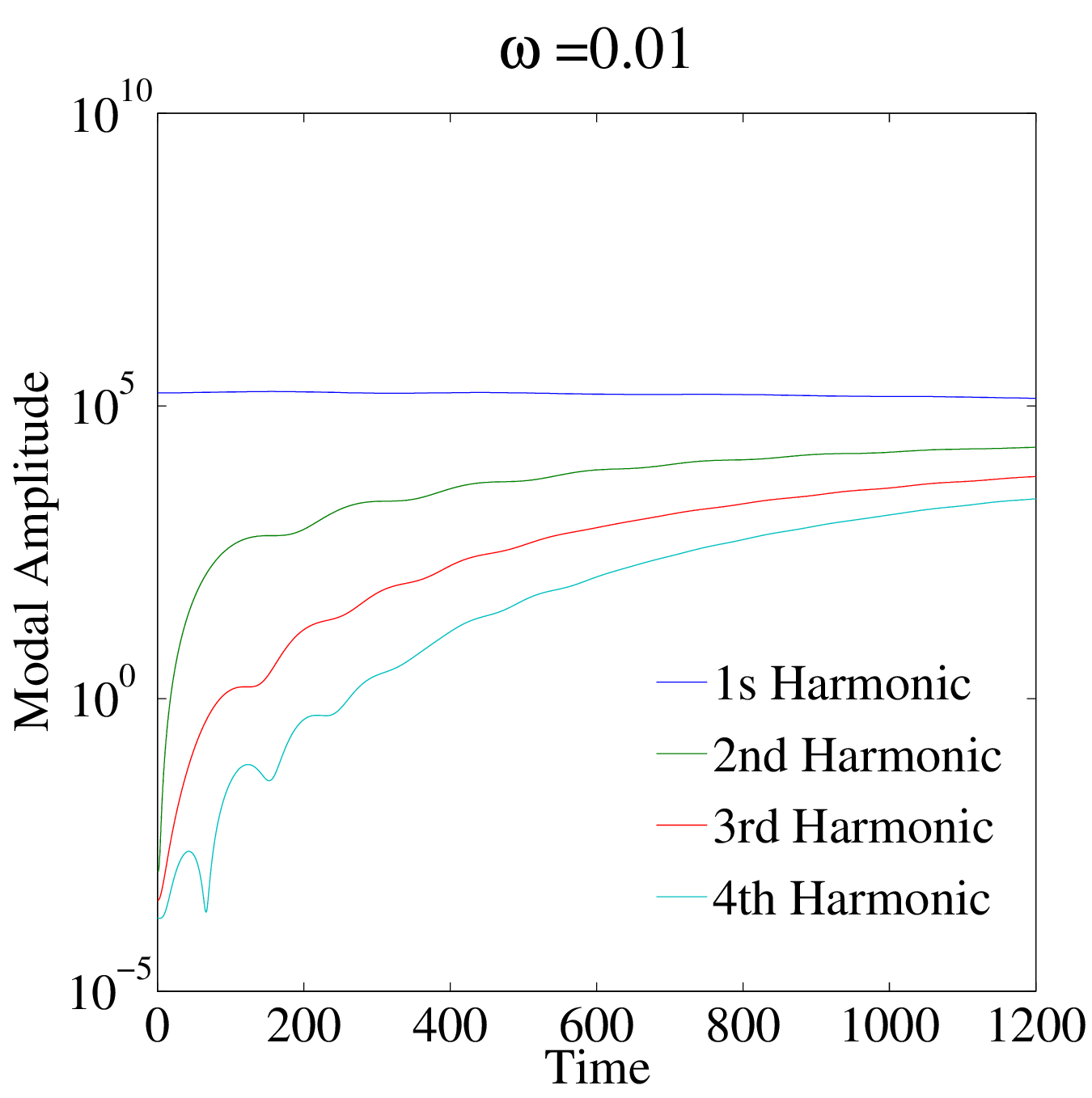}}
\caption{\small\em Upper panel (a): schematic view of the numerical experiments. Here $L$ is the length of the computational domain, $d$ is the uniform water depth, $a_0$ is the incoming wave amplitude and $\lambda$ is its wavelength; lower panel (b): temporal evolution of the first four harmonics of a sinusoidal wave of frequency $\omega = 0.01$ injected in the domain (in the absence of the right wall).}
\label{fig1}
\end{figure}

We consider a two-dimensional wave tank with a flat impermeable bottom of  uniform depth $d = \mathrm{const}$, filled with an incompressible, inviscid fluid (see Figure~\ref{fig1}). The Cartesian coordinate system $Oxy$ is chosen such that the $y-$axis points vertically upwards and the horizontal $x-$axis coincides with the undisturbed water level $y = 0$. The free surface elevation with respect to the still water level is denoted by $y = \eta(x,t)$ and hence, the total water depth is given by $h(x,t) = d + \eta(x,t)$. Denoting the depth-averaged horizontal velociy by $u(x,t)$, the \acs{SGN} system reads \cite{Lannes2009, Dias2010, Clamond2009}:
\begin{eqnarray}\label{eq:serre1}
  h_t + (hu)_x &=& 0, \\
  u_t + \bigl(\half u^2 + gh\bigr)_x &=& \third h^{-1}\Bigl[h^3\bigl(u_{xt} + uu_{xx}
- u_x^2\bigr)\Bigr]_x, \label{eq:serre2}
\end{eqnarray}
where $g$ is the acceleration due to gravity.

The \acs{SGN} system possesses Hamiltonian and Lagrangian structures \cite{Li2002,
Clamond2009} and conservation laws for mass, momentum, potential vorticity and energy \cite{Li2002, Dutykh2011a}. From a more physical perspective the \acs{SGN} model combines strong nonlinear effects with some dispersion that approximates well the full water wave dynamics. This model has been previously validated by extensive comparisons with experimental data for wave propagation and run-up \cite{ChazelLannes2010, Tissier2011, Carter2011, Dutykh2011a}.

One of the most important questions in water wave theory is the understanding of wave interactions and reflections \cite{Linton2001, Berger2003, Clamond2006} and the interaction of solitary waves has also been extensively studied  \cite{Zabusky1965, Bona1980, Fenton1982, CGHHS}. By using symmetry arguments, one can show that the head-on collision of two equal solitary waves is equivalent to the solitary wave/wall interaction in the absence of viscous effects.

The accurate determination of the maximum wave height on a wall is of primary importance for applications. Several analytical predictions for periodic or solitary wave run-up $\R_{\max}$ in terms of the dimensionless wave amplitude $\a = {a_0}/{d}$ have been developed: linear theory \cite{Mei1989} $\R_{\max}/d = 2\a$; third-order theory \cite{Su1980} $\R_{\max}/d = 2\a + 1/2\a^2 + 3/4\a^3$, and nonlinear shallow water theory \cite{Mirchina1984} $\R_{\max}/d = 4\bigl(1 + \a - \sqrt{1 + \a}\bigr) = 2\a +1/2\a^2 - 1/4\a^3 + \O(\a^4)$.

These results have been confirmed in previous experimental \cite{Maxworthy1976}, theoretical \cite{Byatt-Smith1988} and numerical \cite{Fenton1982, CGHHS} studies. All these theories agree on the fact that the wave height on the wall is two times the incident wave amplitude plus higher order corrections. This conclusion provides a theoretical justification for the use of a wave height such as $H_{\nicefrac{1}{3}}$ in the design wave definition.

\section{Numerical study}

From a practical point of view, however, the reasoning presented above contains at least one serious flaw --- in real-world conditions, waves seldom come isolated but rather as groups. In this Section, we show numerically how simple wave groups can produce much higher run-ups than expected from the existing theoretical predictions.

Our numerical wave periods cover a range between $20$ \s{} and $1100$ \s{} (i.e. for $6 < d < 30$ \m{} and $g = 9.81$ \m/\s$^2$), from long swells to tsunami waves.  Extreme run-ups are obtained for wave periods which are in between swell periods and tsunamis periods, corresponding possibly to tsunamis generated by underwater landslides. Moreover, we take as initial conditions waves which are not exact solutions to the equations. Naturally they deform as they evolve towards the vertical wall. This deformation is reminiscent of the transformation of waves over sloping bathymetries.

\subsection{Numerical scheme and set-up}

Let us consider a flat channel of constant depth $d$ and length $L$. This channel is bounded on the right by a rigid vertical wall and by a wavemaker on the left (see Figure~\ref{fig1}). We use dimensionless variables where lengths are normalized with $d$, speeds with $\sqrt{gd}$ and time with $\sqrt{{d}/{g}}$. This scaling is equivalent to setting $g = 1$ \m/\s$^2$, $d = 1$ \m{} in the governing equations \eqref{eq:serre1} and \eqref{eq:serre2}.

In order to solve numerically the \acs{SGN} equations we use the high-order finite-volume scheme described in \cite{Dutykh2011a}. This scheme has been successfully validated against analytical solutions and experimental data \cite{Hammack2004}. For the time integration we use the classical fourth-order Runge--Kutta scheme \cite{Shampine1994, Shampine1997}. The computational domain is divided into equal intervals (i.e. control volumes) such that we have $N = 1000$ control volumes per wavelength. The convergence study showed that this grid provides a good trade-off between accuracy and overall computational time. Note that the extreme run-up values reported in this study can slightly increase under mesh refinement, which decreases the effect of numerical dissipation. All simulations start with the rest state $\eta(x,t=0) \equiv 0$, $\quad u(x,t=0) \equiv 0$.

The wavemaker generates a monochromatic incident wave:
\begin{equation}
  \eta(x=0,t) = \eta_0(t) = a_0\sin(\omega t)\H(T-t),
\end{equation}
where the amplitude $a_0 = 0.05$, $\omega \in [0.01,\; 0.25]$ and $\H(t)$ is the Heaviside function and $c_s$ is the wave speed $c_s= \sqrt{g(d+a_0)}$ \cite{Dutykh2011a}. An important remark should be made on the initialization of the problem: sometimes spurious high-frequency standing waves are generated when numerical simulations of nonlinear progressive waves  are initialized using linear waves (in particular for deep water waves). In such cases the problem should be initialized by suppressing the spurious generation of standing waves. However, a nonlinear simulation can be initialized with simple linear waves if the runtime is long enough to adjust the wave shape \cite{Dommermuth2000}. As can be seen in the lower panel of Figure~\ref{fig1} the amplitude of the first harmonic remains constant during the propagation while the amplitudes of the higher harmonics tend to increase until a constant value is reached, as illustrated in \cite{Dommermuth2000}. For this reason no adjustment is required in our case.

We generate only a finite number $N_w$ of waves with period $T_0 = {2\pi}/{\omega}$ and thus, the wave generation time $T$ is defined as $T = N_wT_0$. The monochromatic deviation of the free surface at the left boundary is then propagated towards the right wall under the \acs{SGN} dynamics. The length $L$ of the computational domain and the final simulation time $T_f$ are chosen adaptively in order to allow all important interactions and to prevent any kind of reflections with the left generating boundary:
\begin{equation*}
  L = (N_w + \half)\lambda, \quad T_f = \frac{L}{\sqrt{g(d+a_0)}} + T,
\end{equation*}
$\lambda$ being the wavelength corresponding to the frequency $\omega$.

\subsection{Numerical results}

\begin{figure}
\noindent
\centering{%
\includegraphics[width=0.99\textwidth]{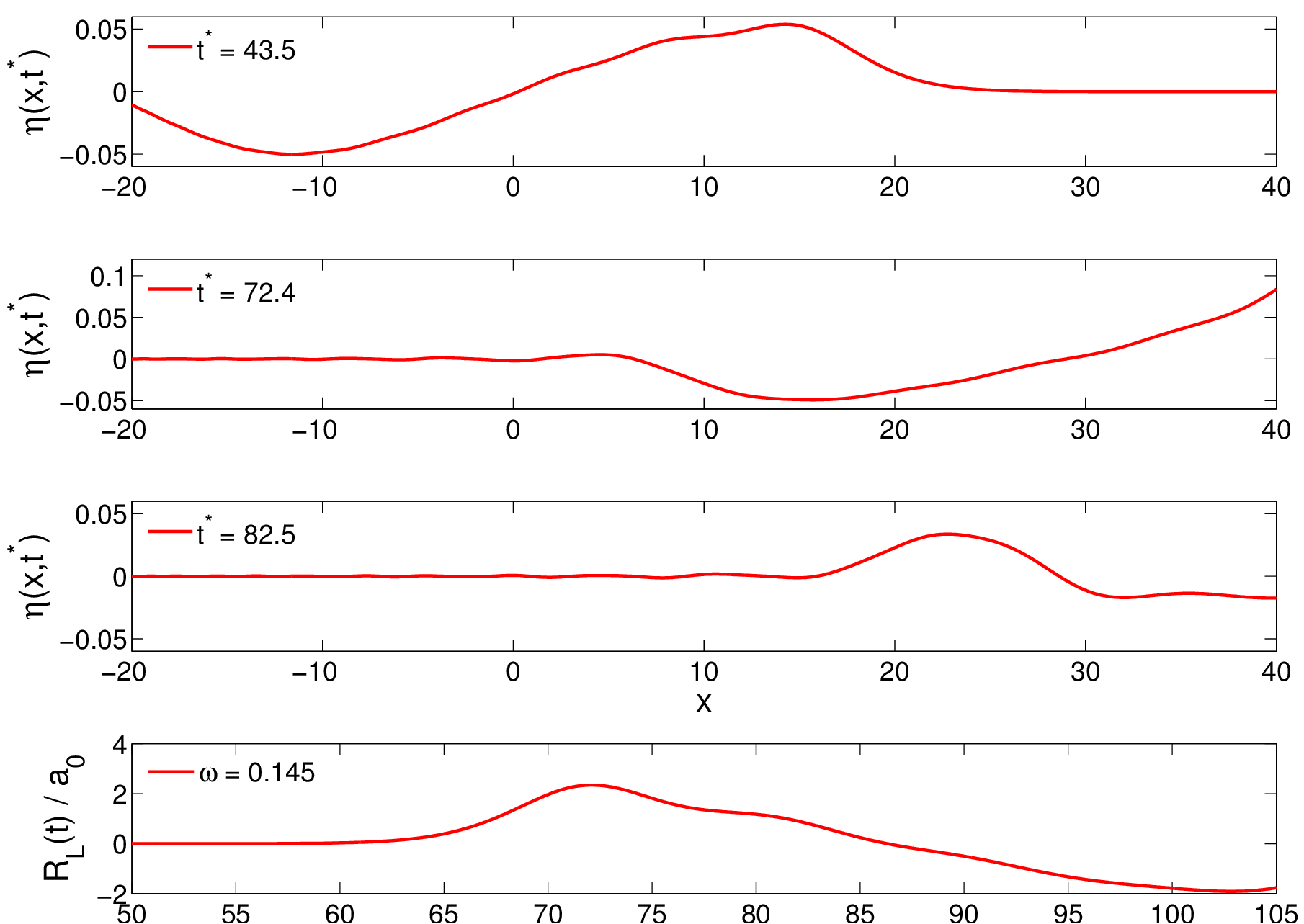}}
\caption{\small\em Time evolution of the free surface elevation as a function of space, at three different times $t^{*}$ (first three top panels) reported on the figure. The figure refers to the single-wave case with frequency $\omega = 0.145$. The lower panel reports the maximal elevation at the wall $\nicefrac{\R_L}{a_0}$ as a function of time.}
\label{fig:snap1}
\end{figure}

We begin our numerical experiments by considering a single sinusoidal wave interacting with the solid wall. In Figure~\ref{fig:snap1} we show three snapshots of the single wave evolution at three different times, i. e. before reaching the wall, during the impact and right after the reflection. The initial sinusoidal wave undergoes steepening during its propagation. The run-up on the vertical wall is shown on the lower panel of Figure~\ref{fig:snap1}. The maximal dimensionless wave elevation $\R_{\max} \simeq 0.10245$ on the wall reaches roughly twice the incident wave amplitude $a_0 = 0.05$ (at $t \simeq 70$). This result is in good agreement with previous numerical studies on the solitary waves interactions \cite{Cooker1997, Pelinovsky1999, Chambarel2009} even if the incident shape is not exactly the same. The maximal relative run-up ${\R_{\max}}/{a_0} \simeq 2.34$ is achieved for $\omega_{\max} = 0.145$. The value of $\R_{\max}$ is slowly decreasing for $\omega > \omega_{\max}$.

\begin{figure} 
\noindent
\centering{%
\includegraphics[width=0.99\textwidth]{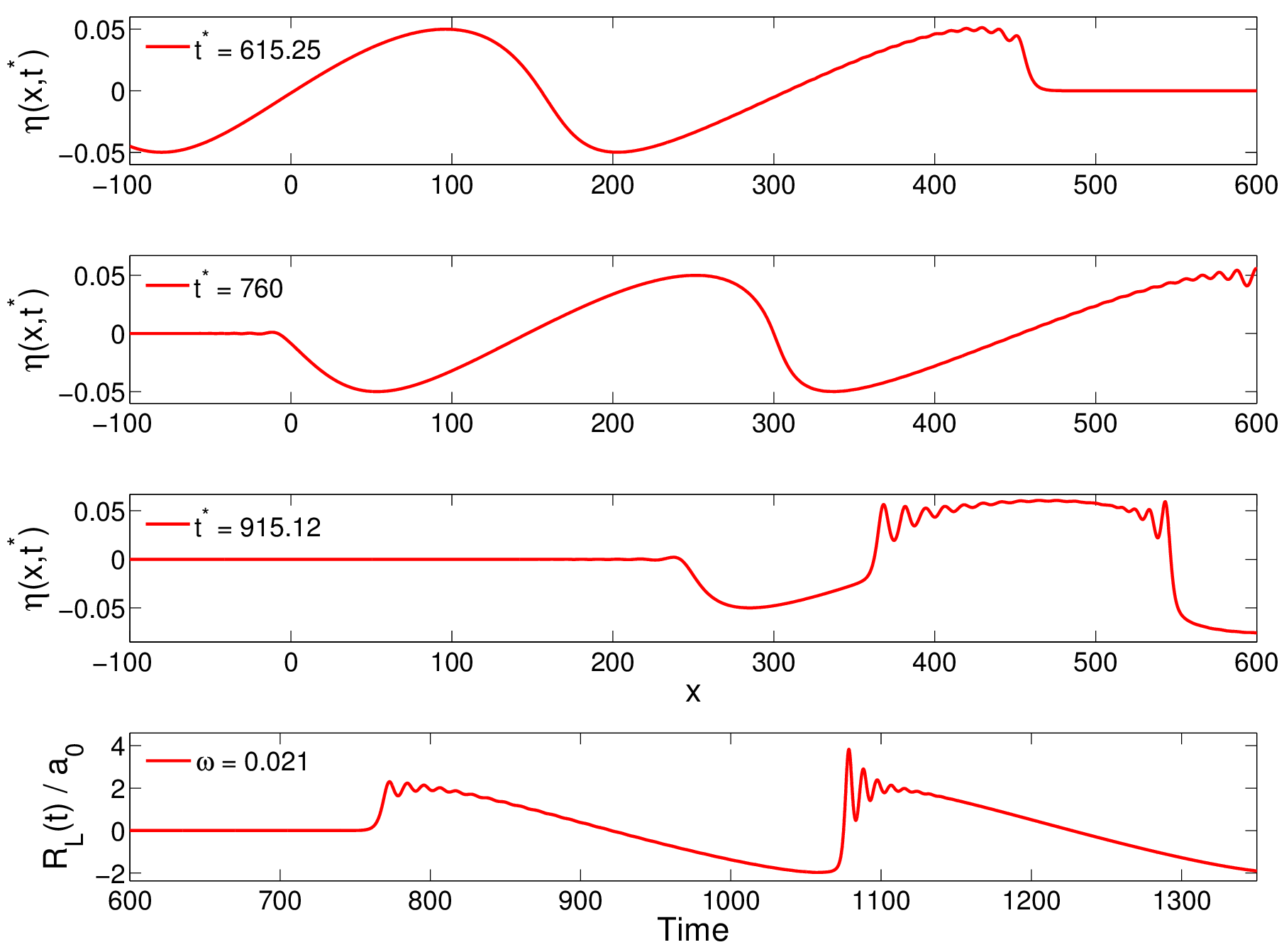}}
\caption{\small\em Time evolution of the free surface elevation as a function of space, at three different times $t^{*}$ (first three top panels) reported on the figure. The figure refers to the two-wave case with frequency $\omega = 0.021$. The lower panel reports the maximal elevation at the wall $\nicefrac{\R_L}{a_0}$ as a function of time.}
\label{fig:snap2}
\end{figure}

When two waves are injected into the domain, the dynamics is similar to the single wave case. However, with two waves the nonlinear effects become even more apparent (see Figure~\ref{fig:snap2}).

\begin{figure}
\noindent
\centering{%
\includegraphics[width=0.95\textwidth]{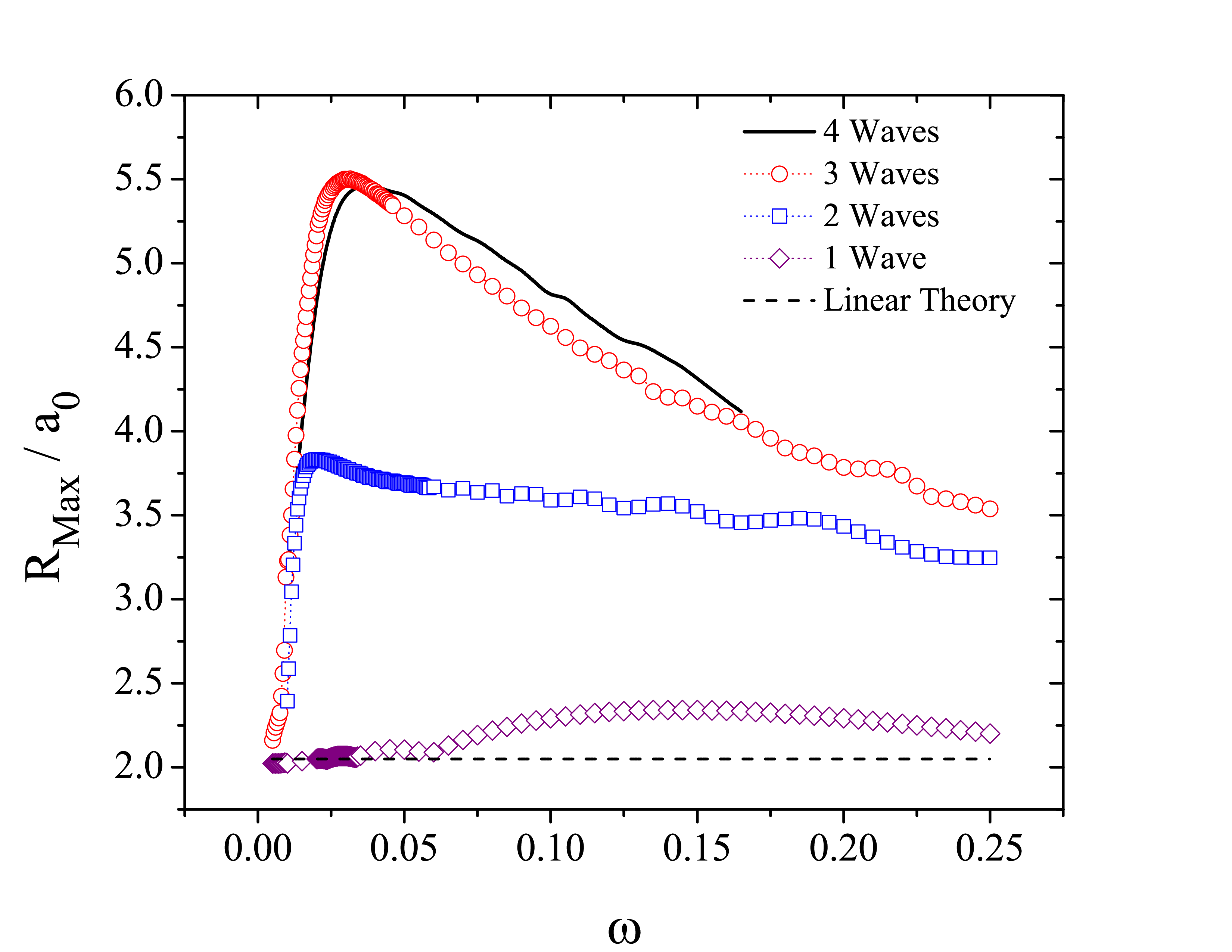}}
\caption{\small\em Maximum wave run-up $\R_{\max}/a_0$ on the right vertical wall as a function of incoming wave frequency for different numbers of incident pulses: 1 (squares), 2 (circles), 3 (triangles) and 4 (solid line). The dashed line represents the linear limit where ${\R_{\max}}/{a_0} \equiv 2$.}
\label{fig:rvsom}
\end{figure}

In a certain range of wave periods ($\omega \in (0.01,\, 0.05)$), the so-called \emph{dispersive shock waves} are formed \cite{Wei1995, Tissier2011}. This particular type of solution has been extensively studied theoretically and numerically in \cite{El2006}. When the second wave impinges on the first reflected wave, a dispersive shock wave forms and propagates towards the wall. As shown in the last panel of Figure~\ref{fig:snap2} the maximal amplification is achieved when the second wave hits the wall due to nonlinear interactions between two counter-propagating waves. However, we underline that with only two waves one can achieve a maximal run-up on the wall $\R_{\max}$ of almost four incident wave amplitudes $a_0$:
\begin{equation}
  \nicefrac{\R_{\max}}{a_0} \simeq 3.8, \quad \mbox{ for } \quad \omega = 0.021.
\end{equation}
Such high run-up values are possible due to the energy transfer between the first reflected wave and the second incoming wave. The dependence of the maximal run-up $\R_{\max}$ on the incident wave frequency $\omega$ and the number $N_w$ of incident waves is shown in Figure~\ref{fig:rvsom}. One can see from this figure that the optimal energy transfer due to dispersive shocks happens for three incident waves (see Figure~\ref{fig:snap3}). In this case the maximal run-up is observed around $\omega_{\max} = 0.035$ and the amplification is equal to $\nicefrac{R_{\max}}{a_0} \simeq 5.43$. However, the energy transfer process is saturated for three waves.

\begin{figure}
\noindent
\centering{%
\includegraphics[width=0.99\textwidth]{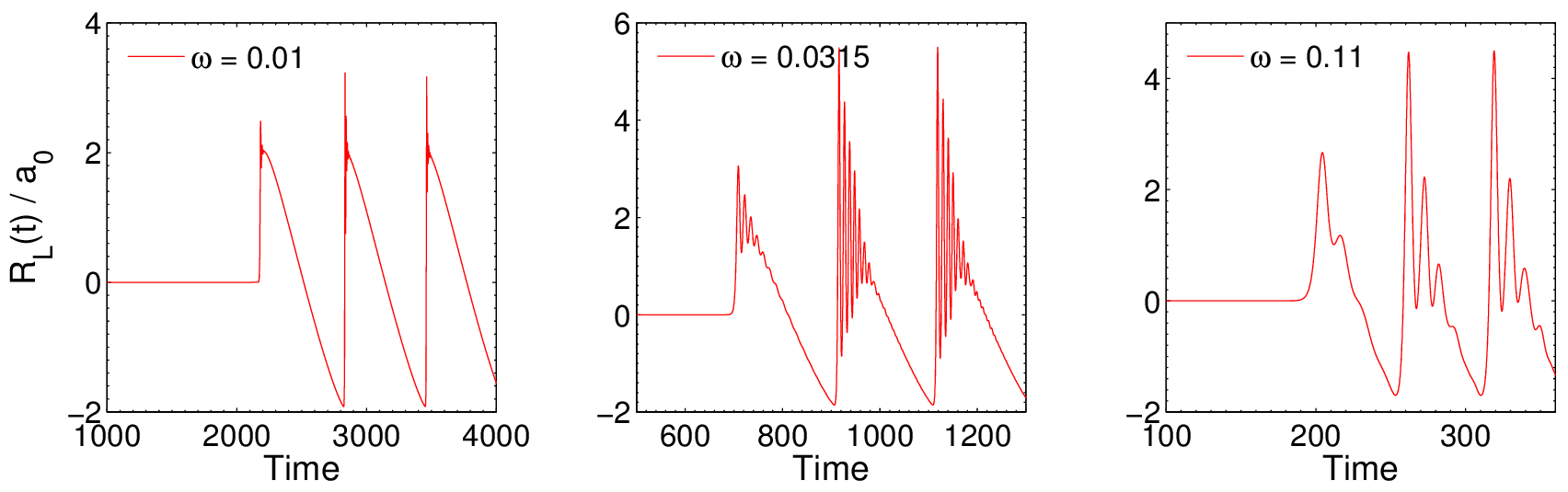}}
\caption{\small\em Time evolution of the wave run-up on the vertical wall for the three incident wave case recorded for several values of the incoming frequency $\omega$. The maximum run-up is achieved for $\omega_{\max} \approx 0.0315$.}
\label{fig:wall}
\end{figure}

\begin{figure}
\noindent
\centering{%
\includegraphics[width=0.99\textwidth]{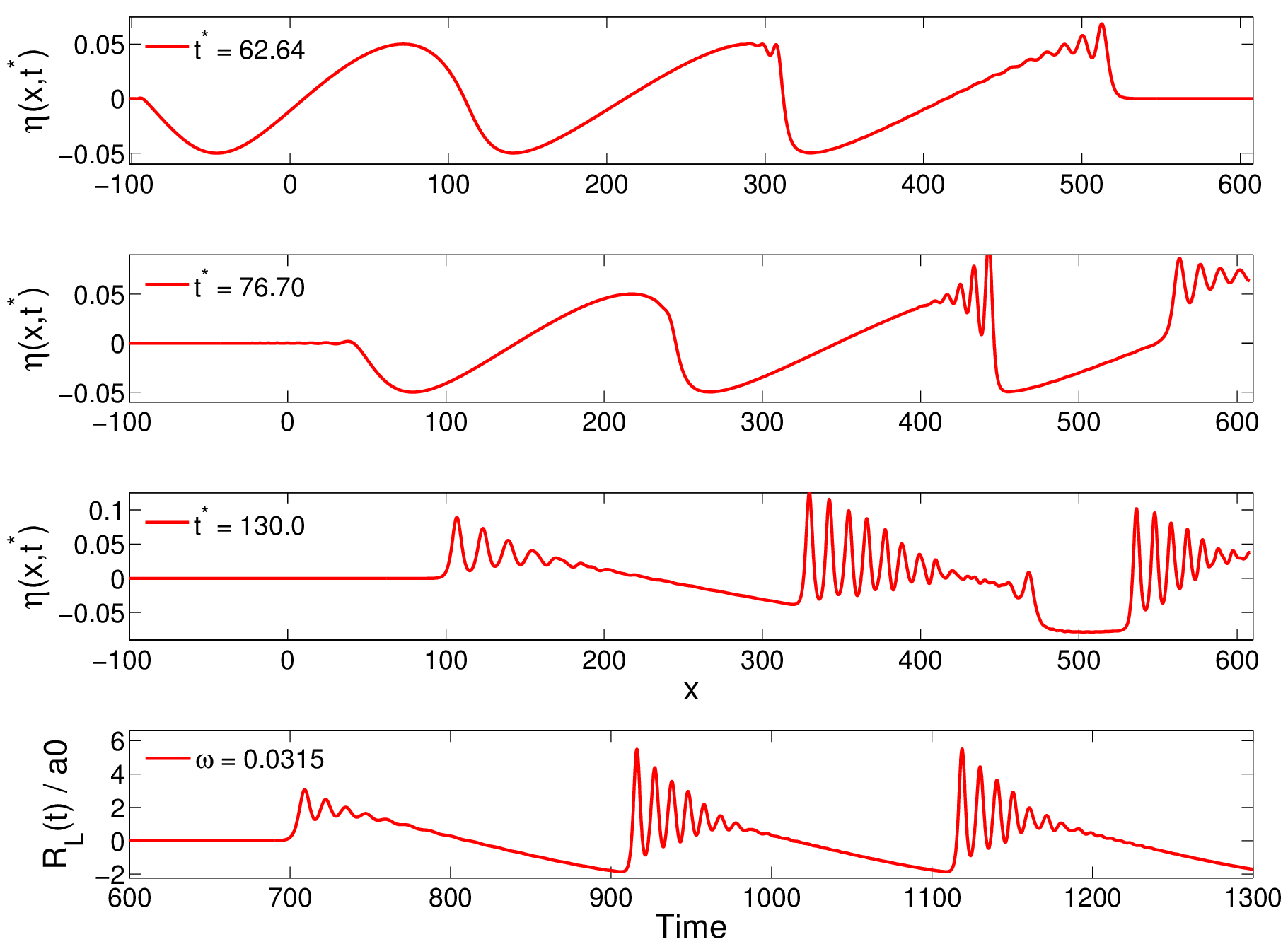}}
\caption{\small\em Time evolution of the free surface elevation as a function of space, at three different times $t^{*}$ (first three top panels) reported on the figure. The figure refers to the three-wave case with frequency $\omega = 0.0315$. The lower panel reports the maximal elevation at the wall ${\R_L}/{a_0}$ as a function of time.}
\label{fig:snap3}
\end{figure}

\begin{figure}
\centering
\subfigure{\includegraphics[width=0.33\textwidth]{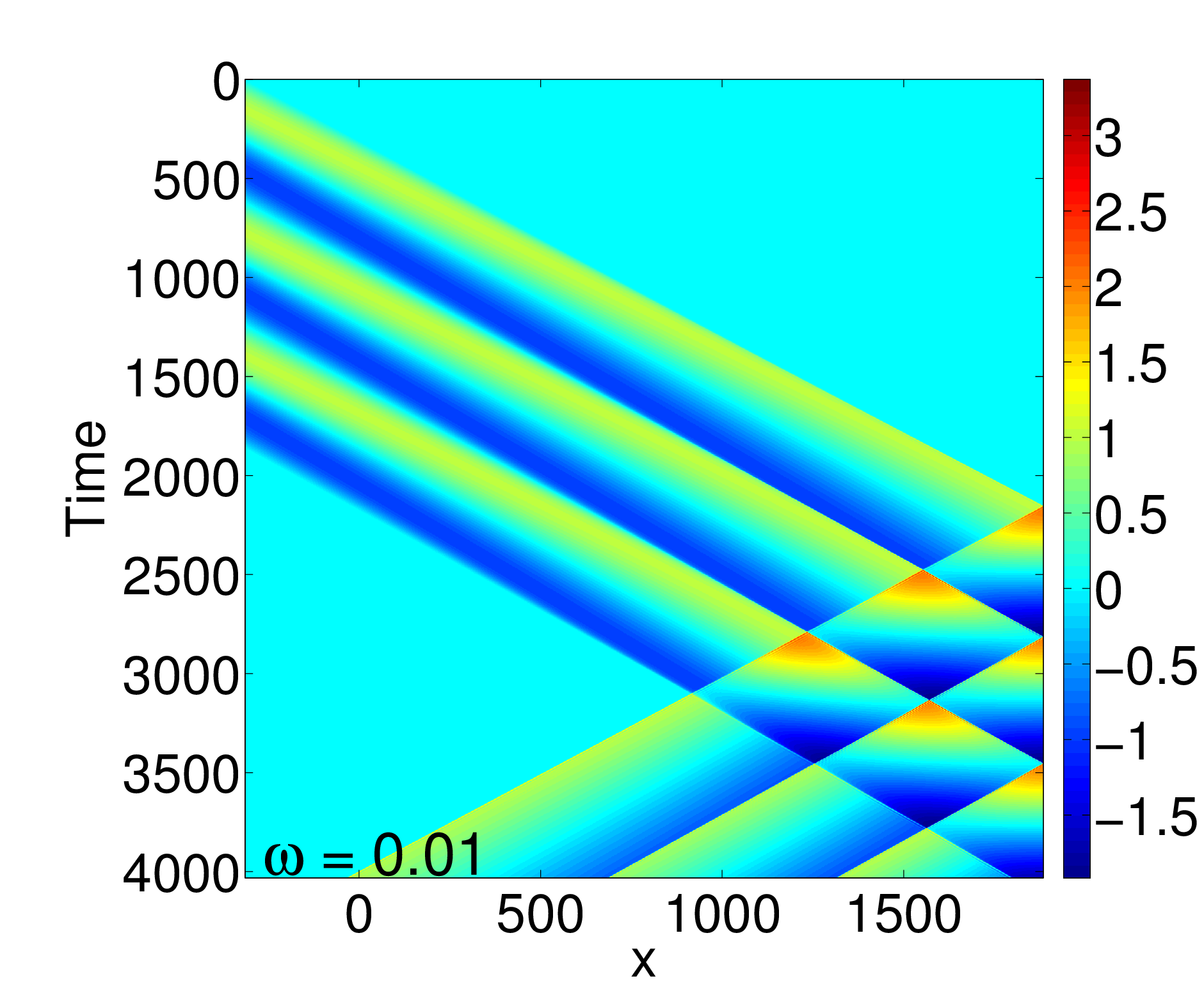}}
\subfigure{\includegraphics[width=0.33\textwidth]{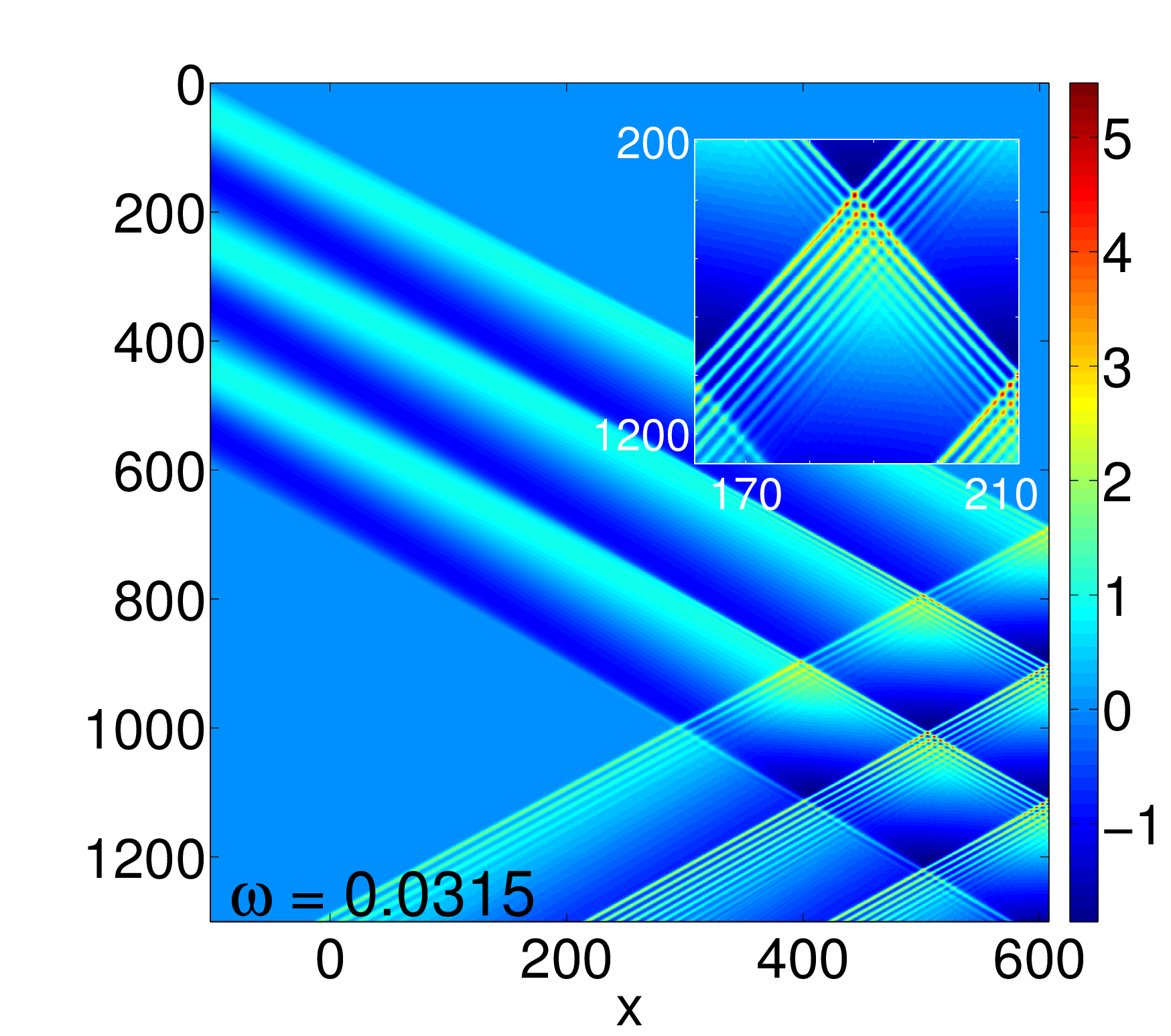}}
\subfigure{\includegraphics[width=0.32\textwidth]{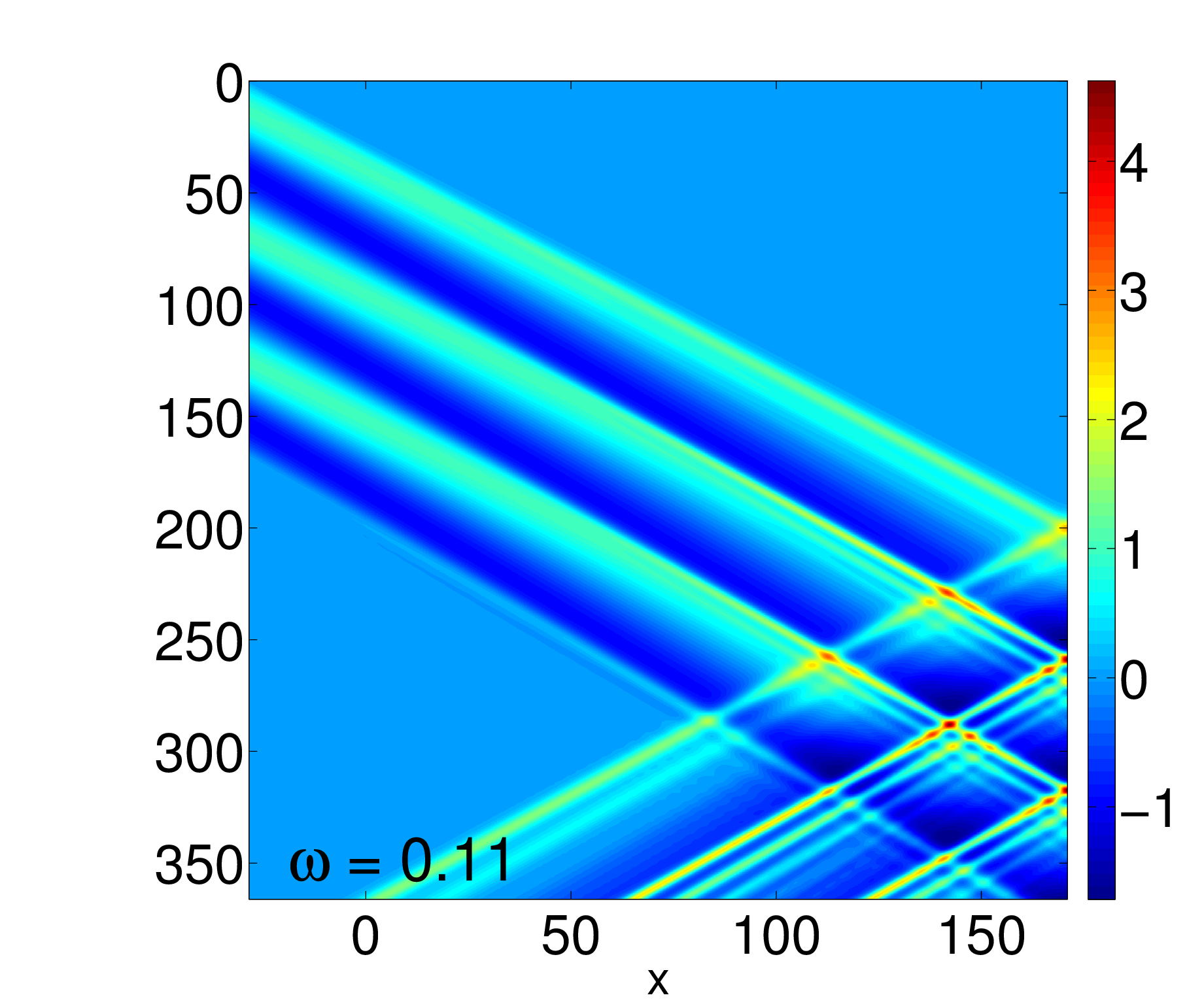}}
\caption{\small\em Space-time evolution plots for the three incident wave case shown for three particular values of the wave frequency $\omega$.}
\label{fig:spacetime}
\end{figure}

We performed similar computations with four incident waves (also shown on Figure~\ref{fig:rvsom}) and the maximal run-up is not higher than with three waves. Consequently, we focus now only on the \emph{optimal} three-wave case. The three regimes (hyperbolic, equilibrium and dispersive) are illustrated on Figure~\ref{fig:spacetime}, where we show the space-time dynamics of the three-wave system. The left panel shows the hyperbolic regime. On the central panel strong dispersive shocks can be observed, while on the right panel the dynamics is smoothed by the dispersion. In the last case the amplification is mainly produced by the linear superposition of the incident and reflected waves. The reflection and interaction are clearly observed by smooth secondary peaks in the space time plots (see Figure~\ref{fig:spacetime}, but also Figure~\ref{fig:wall}).

The wave interactions described above strongly depend on the frequency $\omega$ of the impinging waves as can be seen in Figure~\ref{fig:wall}, where we show the wave records on the wall for several values of the frequency $\omega$. As the wave frequency increases, the wavelength shortens and the dispersive effects become gradually more important. Around $\omega_{\max}$ the dispersive effects are balanced with nonlinearities to produce the most pronounced dispersive shock waves. Starting from $\omega \simeq 0.11$ we enter into the dispersive regimes where the waves become regularized.

\section{Conclusions and perspectives}

In the present study we investigated numerically the interaction between a wave and a vertical wall in the framework of the \acf{SGN} equations. These equations combine strong nonlinear and weak dispersive effects. We explored the whole range of wavelengths from the hyperbolic regime (including shock waves) to smooth dispersive waves ($\omega \gtrsim 0.1$). More importantly, we showed that the wave run-up on the wall is strongly dependent on the incident wave frequency, the dependence not being monotonic. In particular, there is a fixed frequency $\omega_{\max}$ which provides the maximum long wave run-up on the vertical wall. The function $\R_{\max}(\omega)$ is monotonically increasing up to $\omega_{\max}$ and monotonically decreasing for $\omega > \omega_{\max}$ (at least within the range of considered numerical parameters). The dynamics can be conventionally divided into three main stages:
\begin{enumerate}
  \item Propagation of the wave group towards the wall along with the front steepening and other nonlinear deformations
  \item First wave run-up and its reflection from the wall
  \item Projection of the reflected energy again onto the wall by subsequent incident waves
\end{enumerate}
The maximal observed amplification $\approx 5.5a_0$ is achieved with only three incident waves. For example, on a $10$ $m$ water depth the \emph{critical} period is equal approximatively to $T_{\max} \approx 180$ $s$ ($\omega_{\max} \equiv {2\pi}/{T_{\max}}$). Such a wave period can be generated by small underwater landslides.

The results presented in this study shed light onto extreme wave run-ups on vertical cliffs and similar coastal structures. Moreover, in view of these results, the definition of the \emph{design wave} has to be revisited. Our suggestion would be to take at least $3H_{\nicefrac{1}{3}}$ or even $3H_{\nicefrac{1}{10}}$. The present results also shed some new light on the mysterious accumulations of large boulders on cliff tops up to 50 \m{} high on the deep water coasts, especially on the west coast of Ireland \cite{Hansom2009, Williams2010, O'Brien2013}. The emplacement of these megaclasts is usually attributed to extreme storm waves, but there are also those who believe that tsunamis are the most probable explanation of boulder ridges in these areas \cite{Kelletat2008, Scheffers2009, Scheffers2010}.

In future investigations, more general wave groups have to be studied to unveil their potential for focussing on the walls. We recall that so far we considered only simple idealized monochromatic waves. In addition, we are going to investigate the effect of the  forces exerted by incident waves on vertical obstacles, which can be different from the purely kinematic amplitude focussing presented in this study. In other words, it is not clear whether the highest wave will produce the highest dynamic pressure spike on the wall. The effect of the wave amplitude is to be investigated as well since all the processes under consideration are highly nonlinear. Some theoretical explanation of these phenomena is also desirable. However, the difficulty is rather high again because of important nonlinearities mentioned hereinabove. We claim that no linear theory is sufficient to provide a satisfactory explanation.

\subsection*{Acknowledgments}
\addcontentsline{toc}{section}{Acknowledgments}

This work was funded by ERC under the research project ERC-2011-AdG 290562-MULTIWAVE and SFI under Grant Number SFI/12/ERC/E2227. The authors wish to acknowledge the SFI/HEA Irish Centre for High-End Computing (ICHEC) for the provision of computational facilities and support under the project ``\textit{Numerical simulation of the extreme wave run-up on vertical barriers in coastal areas}''. We would like also to thank Aliz\'ee \textsc{Dubois} who obtained the first preliminary results on this subject during her visit to University College Dublin in May 2012.

\addcontentsline{toc}{section}{References}
\bibliographystyle{abbrv}
\bibliography{biblio}

\end{document}